# Recent Advances in Bubble-Assisted Liquid Hole-Multipliers in Liquid Xenon

E. Erdal,[a] L. Arazi,[b,*] A. Tesi [a], A. Roy,[a] S. Shchemelinin,[a] D. Vartsky,[a] and A. Breskin [a]

[a] *Department of Particle Physics and Astrophysics, Weizmann Institute of Science, Rehovot 7610001, Israel*

[b] *Nuclear Engineering Unit, Faculty of Engineering Sciences, Ben-Gurion University of the Negev, P.O. Box 653, Beer-Sheva 1084548, Israel*

  *E-mail*: larazi@bgu.ac.il

ABSTRACT: We report on recent advances in the operation of bubble-assisted Liquid Hole-Multipliers (LHM). By confining a vapor bubble under or adjacent to a perforated electrode immersed in liquid xenon, we could record both radiation-induced ionization electrons and primary scintillation photons in the noble liquid. Four types of LHM electrodes were investigated: a THGEM, standard double-conical GEM, 50 µm-thick single-conical GEM (SC-GEM) and 125 µm-thick SC-GEM – all coated with CsI photocathodes. The 125 µm-thick SC-GEM provided the highest electroluminescence (EL) yields, up to ~400 photons per electron over $4\pi$, with an RMS pulse-height resolution reaching 5.5% for events comprising ~7000 primary electrons. Applying a high transfer field across the bubble, the EL yield was further increased by a factor of ~5. The feasibility of a vertical-mode LHM, with the bubble confined between two vertical electrodes, and the operation of a two-stage LHM configuration were demonstrated for the first time. We combine electrostatic simulations with observed signals to draw conclusions regarding the location of the liquid-gas interface and suggest an explanation for the observed differences in EL yield between the investigated electrodes.

---

[*] Corresponding author.

# Contents

## 1. Introduction

The bubble-assisted liquid hole-multiplier (LHM) is a recent concept, proposed for the combined detection of ionization electrons and primary scintillation photons generated along charged-particle tracks in noble liquids [1-5]. A conceptual scheme of an LHM module is depicted in Figure 1. It consists of a perforated electrode, e.g., a gas electron multiplier (GEM) [6], or a thick gas electron multiplier (THGEM) [7] immersed inside the liquid, with a bubble of the noble gas trapped underneath. The top surface of the electrode is optionally coated with a VUV-sensitive photocathode such as cesium iodide (CsI). Heating elements (such as a plane of heating wires) below the electrode are used to generate the bubble on demand. Once formed, the bubble fills the space below the electrode and remains stable as long as the system is in a thermodynamic steady state, with buoyancy pushing it upward and surrounding walls confining it from its sides. The top and bottom surfaces of the electrode, as well as the heating wires, are held at different potentials, creating a dipole-like field in the electrode's holes and a transfer field between the electrode bottom and the wire plane. The drift field above the electrode is set by the potential difference between its top surface and a distant drift electrode (cathode plane) above. Particle interactions in the liquid lead to a prompt scintillation signal (S1) and to the release of ionization electrons. These drift towards the LHM and are focused by the local field into its holes. Once they cross the liquid-gas interface into the bubble, they induce electroluminescence (EL) light (S2) in the gas, in the high-field region close to the bottom of the hole. Similarly, S1 photons impinging on the photocathode release photoelectrons (PEs); these are focused into the LHM holes, inducing an EL signal (S1') tens to hundreds of nanoseconds after S1 (depending on the electrode used) and typically long before S2 (from microseconds to milliseconds, depending on the drift distance). The S1' and S2 EL signals can be recorded by a position-sensitive photon detector (e.g., a SiPM array) located below the bubble, allowing for accurate reconstruction of their 2D location.

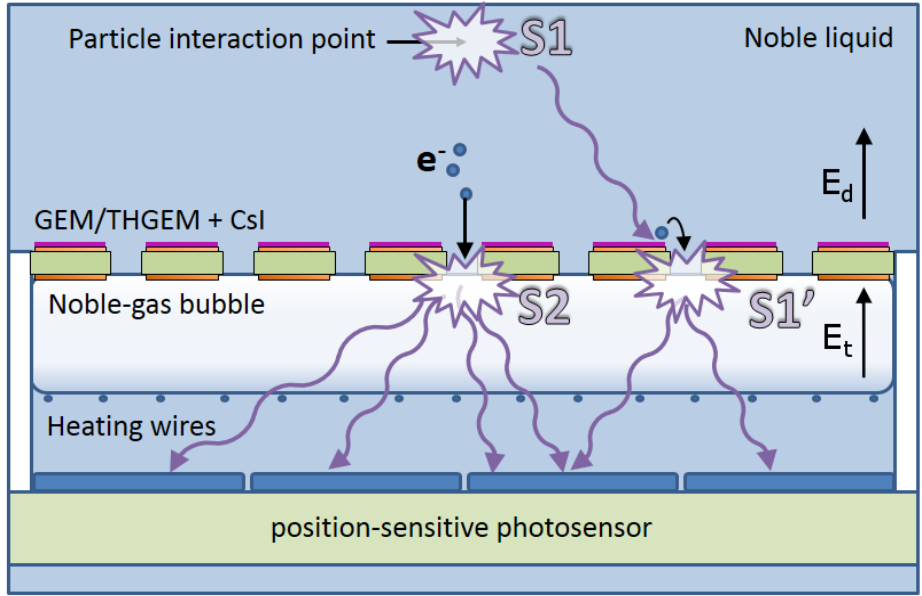

**Figure 1**: Conceptual scheme for an LHM module, comprising a perforated electrode (e.g., GEM or THGEM) coated on top with a photocathode (e.g., CsI), a grid of heating wires for forming the bubble and a position-sensitive photon detector below. The bubble is supported by buoyancy against the bottom of the electrode. Ionization electrons focused into the holes create EL light (S2) once they cross the liquid-gas interface into the bubble. Primary scintillation (S1) photons impinging on the photocathode release photoelectrons which are focused into the holes and create similar EL signals (S1'). The lateral coordinates of the S2 and S1' signals are reconstructed by the position-sensitive photon detector.

The LHM concept was first suggested in [1] as a possible single-phase detector element for future multi-ton noble-liquid time projection chambers (TPCs) for rare-event experiments. The bubble-assisted LHM is presently considered as a potential building block in the design of the DARWIN ~50 ton liquid-xenon dark-matter observatory [8, 9].

Previous works performed with a configuration similar to that shown in Figure 1 demonstrated the idea, using GEM and THGEM electrodes immersed in liquid xenon (LXe), with alpha-particle tracks providing primary scintillation light and ionization electrons. It was shown that bubbles can indeed be formed on demand by ohmic heating and that once formed, they remain stable as long as the system stays in a thermodynamic steady state. (So far, complete bubble stability was observed over 10 days - until the experiment was aborted to free the cryostat for further studies [4].) The electrodes used so far had an active diameter of 14 or 20 mm, with bubbles filling a ~35 mm diameter region underneath. The EL yield was estimated to be ~50-100 photons per electron entering the hole [3]. The RMS S2 energy resolution was shown to be 8% for ~10,000 ionization electrons. When coated with CsI, GEM electrodes displayed similar energy resolutions for a few hundred VUV-induced photoelectrons, with a corresponding 4 ns RMS time resolution [4, 5].

In the present work we report on progress made on several aspects of the LHM study. This includes: (1) a comparison between four different types of LHM electrodes (THGEM, standard double-conical GEM, 50 μm-thick single-conical GEM and 125 μm-thick single-conical GEMs; (2) demonstration of major amplification of the EL light yield by setting a strong transfer field across the bubble; (3) first demonstration of a vertical-LHM operation mode, where the bubble is confined between two vertical electrodes, and (4) first demonstration of a cascaded double-LHM structure, with a bubble trapped below each element. The horizontal-LHM data are analyzed with the help of electrostatic simulations, providing insight regarding the location of the bubble interface.

## 2. Experimental setup and methodology

The experiments were conducted in a dedicated LXe cryostat, the Mini Xenon apparatus (MiniX), described in detail in [4]. It comprises a 100 mm-diameter, 100 mm-tall cylindrical LXe volume filled with ~250 ml of LXe. The rest of the volume is equipped with instrumentation and PTFE holders and spacers. The detector assemblies are suspended from the topmost flange. The cryostat has a window which allows viewing the detector from below at 60° with respect to the vertical axis. Xenon liquefaction is done on the $LN_2$-cooled cryostat wall. During operation, LXe is continuously extracted through a tube immersed inside the liquid and circulated through an SAES hot getter at 2 standard liters per minute. The purified Xe gas returns to the vapor phase at the top part of the chamber through a tubular heat exchanger (where heat transfer takes place with the extracted LXe), and liquefies on the chamber wall which is kept at 170 K (for a schematic drawing and additional details see [4]). Before mounting detector assemblies incorporating CsI photocathodes, the cryostat was pumped down to high vacuum using a turbo-molecular pump and a $LN_2$-cooled cold finger to reach a water partial pressure below $10^{-7}$ mbar. Assembly of the detector setups in the cryostat was done under constant Ar flushing, to minimize moisture contamination[1]. Subsequent steps included pumping down the cryostat for at least 2 days, filling Xe gas at 2 bar, circulating the gas through the SAES purifier for 1-2 days, cooling down, further filling under Xe liquefaction, recirculation of the liquid through the SAES purifier for 2 days, and bubble formation by ohmic heating. The duration of gas recirculation prior to bubble formation was chosen based on past observation that the EL yield improves during the first day of purification and then becomes stable. For each investigated configuration, measurements were carried out over several days to confirm that there are no further changes of EL yield at a given voltage. Measurements were conducted at a liquid temperature of ~173 K, corresponding to a vapor pressure of 1.6 bar.

---

[1] Flushing was done using Ar rather than $N_2$, since it is heavier than air and therefore is more effective in preventing air from entering the chamber through the top flange during installation of the inner setup. Since the addition of small amounts of Xe to Ar gas is known to strongly shift the emission wavelength from that of the argon excimer (128 nm) to that of xenon (171 nm) [10], we expect that tiny amounts of Ar in Xe will have negligible effect on the emission spectrum with respect to pure Xe and will therefore have no effect on our measurements.

## 2.1 Horizontal LHM

Four LHM electrodes were investigated in a horizontal configuration (as in Figure 1): a THGEM, a standard 50 μm-thick GEM with bi-conical holes and two single-mask GEMs with conical holes (to which we hereafter refer to as single-conical GEM, or SC-GEM) [11] - 50 and 125 μm-thick. (The bi-conical and single-conical shapes of the GEM and SC-GEM holes result from the etching process employed in their production - from two sides in the case of the GEM and one side in that of the SC-GEM.) The Cu surfaces of all electrodes were Au-plated. The geometrical properties of the investigated electrodes are listed in Table 1, and 3D models of their respective unit cells (used for electrostatic simulations described in Appendix A) are shown in Figure 2. A ~300 nm-thick CsI layer was vacuum-deposited on the top surface of electrodes investigated for UV-photon detection (the SC-GEM was CsI-coated on the side with the smaller holes to maximize the photosensitive area). The CsI-coated electrodes were transferred in dry $N_2$ from the deposition chamber into an $N_2$-filled glove box for assembling the detector setup, which was then quickly transferred (~20 s in ambient air) into the Ar-flushed cryostat. The measured quantum efficiency (QE) value of all photocathodes investigated was in the range of 21-23% at 171 nm (Xe vapor emission wavelength [12]) in vacuum; the relative loss in QE during transfer to the $N_2$-box and to the cryostat was measured to be below 10% (considered to be due to exposure to humidity [13]).

**Table 1:** Specifications of the four electrodes used in this study

|  | THGEM | Standard GEM | Single-conical GEM | Single-conical GEM |
|---|---|---|---|---|
| Insulator | FR4 | polyimide | polyimide | polyimide |
| Thickness | 0.4 mm | 50 μm | 50 μm | 125 μm |
| Hole diameter(s) | 0.3 mm | top/mid/bottom 70/50/70 μm | top/bottom 300/340 μm | top/bottom 300/400 μm |
| Hole pitch | 1 mm | 140 μm | 600 μm | 600 μm |
| Cu thickness | 20 μm | 5 μm | 5 μm | 5 μm |
| Hole rim | 50 μm | -- | -- | -- |

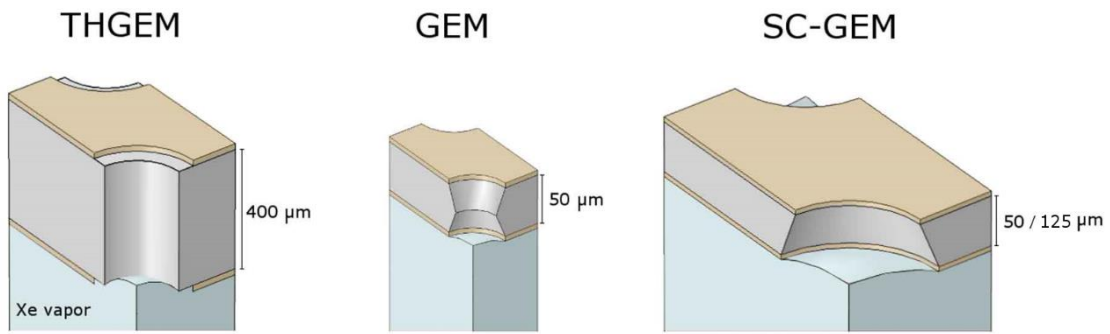

**Figure 2:** 3D models of unit cells of the three LHM electrodes used in the experiments and in the accompanying simulations. The electrodes are immersed in LXe; the region below each electrode (in light blue) is Xe vapor, with a liquid-gas spherical interface penetrating into the holes. Note the different scale for the THGEM compared to the GEM and SC-GEM.

The detector assembly (Figure 3) comprised the CsI-coated LHM electrode, a spectroscopic ~180 Bq, 6 mm diameter $^{241}$Am alpha-particle source located 7.9 mm above it, a field-shaping ring between the electrode and the source and a resistive-wire plane underneath. The field-shaping ring (inner diameter 20 mm) was introduced for creating a fairly uniform field across the central region of the photocathode, without the need for an intermediate mesh (which would have resulted in partial blocking of the drifting ionization electrons, or alternatively require an intense field between the mesh and electrode, thus interfering with photoelectron extraction from the photocathode). The active (perforated and CsI-coated) area of the GEM and the SC-GEM electrodes was 14 mm in diameter; for the THGEM, CsI was deposited on a 20 mm diameter area at the center of the electrode. The resistive-wire plane, located 1.6 mm below each of the LHM electrodes, served two purposes: (1) allow for bubble formation by ohmic heating, and (2) set the value and direction of the transfer field below the electrode (allowing to either pull the electrons through the bubble or push them towards the electrode bottom); as discussed below, the transfer field can also serve to amplify the EL light by a large factor. Once formed, the bubble extended from the electrode bottom to the wires, filling a 35 mm-diameter region, as shown schematically in Figure 3.

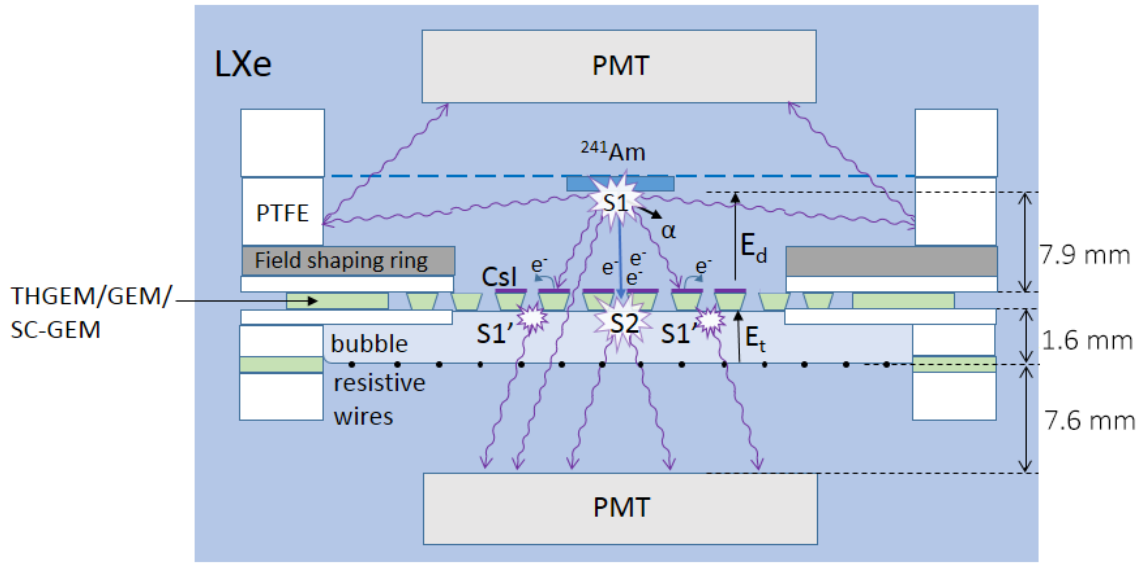

**Figure 3:** Schematic setup for the comparison between the THGEM, GEM and SC-GEM (shown as an example) electrodes as horizontal bubble-assisted LHM elements; dimensions not to scale

Light signals were recorded by two Hamamatsu R8520 photomultiplier tubes (PMTs). The top PMT recorded S1 photons reflected off the PTFE ring and served as the trigger. S1 photons passing through the electrode's holes, as well as photons of the S1' and S2 signals were recorded using the bottom PMT located below the LHM electrode. The signals of the two PMTs were directed to a Tektronix MSO5204B oscilloscope through a Phillips 777 linear amplifier unit (with gain ~1.8). Waveforms were digitized and saved using the oscilloscope and were post-processed using dedicated Matlab scripts.

## 2.2 Vertical LHM

The basic concept of the bubble-assisted LHM relies on the ability to confine a bubble in contact with a hole-electrode such that electrons deposited and drifting in the liquid are efficiently focused

– 5 –

into the bubble. This concept could in principle be realized not only by supporting the bubble underneath a horizontal electrode but also by confining a bubble in a vertical "cage", of which one side consists of a hole-electrode and the other - of a fine mesh or transparent plate.

The vertical LHM setup, shown in Figure 4, consisted of the spectroscopic $^{241}$Am alpha-particle source positioned behind a bubble-confinement cage, comprising PTFE spacers, a GEM and a fine-pitch electroformed Cu mesh (Precision Eforming MC-32; opening: 112 μm). A resistive heating wire was introduced through two small holes at the bottom of one of the PTFE spacers, to form the bubble on demand. The setup was placed inside the liquid volume, where the bubble could be observed at 60° with respect to the vertical axis of the cryostat through the window. In the preliminary experiments reported here, the bubble was formed spontaneously by heat leaks (through the Kapton-insulated wires biasing the setup components). Once formed, it displayed periodic dynamics of abrupt partial shrinking followed by gradual growth over the entire field of view, with a period of ~20 s. EL signals from the bubble were recorded by a PMT which – due to geometrical constraints of the cryostat – had to be placed outside of the vessel at a distance of ~20 cm from the LHM. Light collection in this geometry was low, resulting in poor photon statistics. To obtain sufficiently large S2 EL signals, an intense transfer field was applied across the bubble. The low rate of background events (~1 Hz) in the absence of voltages across the electrodes, helped identifying the ~100 Hz S2 events once voltages were set to pull the electrons through the electrodes to the bubble (consistent with the source activity). Triggering was done on S2 and the PMT signals were digitized and stored by the Tektronix MSO5204B oscilloscope. Post-processing allowed to readily discard 'bad' events where the bubble did not fill the region in front of the source.

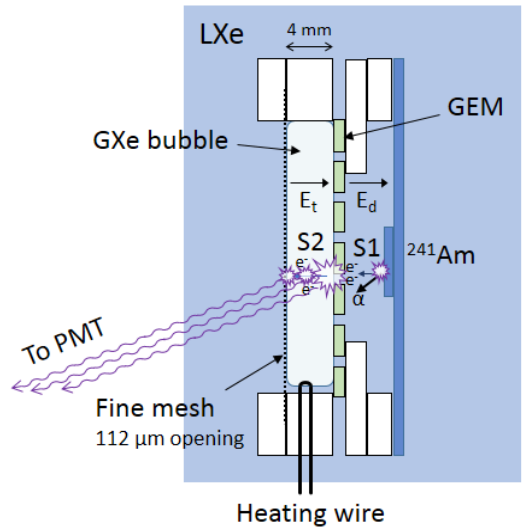

**Figure 4:** The vertical-LHM assembly for the confinement of a bubble between a vertical GEM electrode and a fine-pitch mesh (not to scale). PTFE spacers define a closed "cage" between the electrode and the mesh. A thin resistive heating wire is introduced through small apertures into the cage, to generate the bubble. To overcome the low light collection efficiency, S2 EL signals were amplified in the transfer gap by applying an intense transfer field across the bubble.

## 2.3 Double-stage LHM

We present here a first proof-of-principle of the cascaded LHM concept [1]. We investigated a double-stage detector, shown in Figure 5, incorporating two LHM elements, each having a bubble underneath to generate EL signals. In the spirit of the original LHM idea, radiation-induced EL photons from the first stage impinge on a VUV-sensitive photocathode deposited on the second hole-electrode – generating additional EL signals. This concept is similar to the "photon-assisted" cascaded detector developed for ion blocking in gas-avalanche detectors [14-16]. We report here on preliminary results obtained with a double-stage LHM comprising two SC-GEM elements, with a CsI photocathode deposited on top of the second one.

The cascaded LHM setup (Figure 5) comprised two SC-GEM electrodes (different from the ones described in table 1: the top and bottom hole diameters were 150 μm and 190 μm respectively, the hole spacing was 300 μm and Kapton thickness 50 μm); only the bottom SC-GEM was coated with CsI. Bubbles were formed underneath, generated by resistive wires located below each electrode. We refer to the inter-electrode gap and the gap between the second electrode to the wire-plane below as transfer gaps 1 and 2, respectively, with their associated nominal transfer fields $E_{t1}$ and $E_{t2}$. The $^{241}$Am spectroscopic alpha-particle source was located 5 mm above the first LHM electrode, and triggering on reflected S1 light was done by the top PMT; all EL signals were recorded by the bottom PMT, placed below the second SC-GEM electrode.

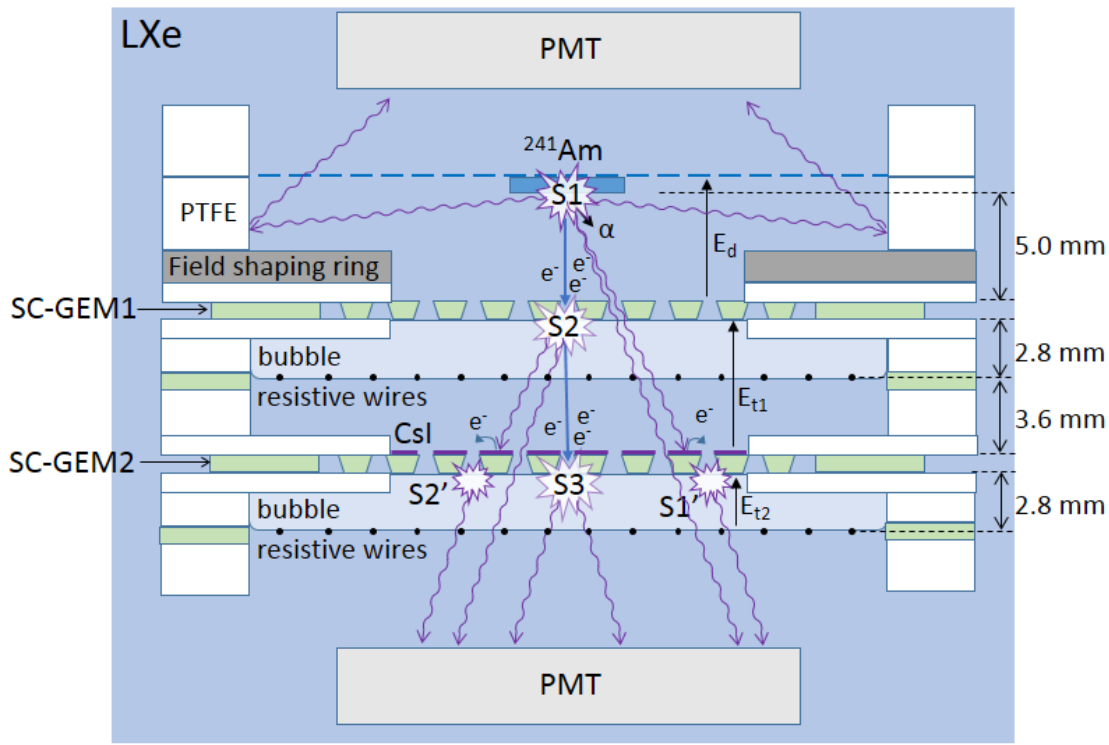

**Figure 5**: Schematic drawing of the double-stage bubble-assisted LHM setup. Two SC-GEMs have heating wire planes underneath, generating independent bubbles under each hole-electrode. The bottom PMT records prompt scintillation photons (S1), EL generated by photoelectrons on second stage (S1' and S2'), EL generated by ionization electrons on first stage (S2) and EL generated by ionization electrons on second stage (S3).

As depicted in Figure 5, following alpha emission from the source into the liquid, a fraction of the S1 photons are reflected off the PTFE walls, reaching the top PMT. A small fraction of the S1 light penetrates through both SC-GEMs, reaching the bottom PMT. Another fraction reaches the bottom, CsI-coated, SC-GEM2 electrode; the resulting emitted photoelectrons generate an S1' EL signal. The radiation-induced ionization electrons are focused into the holes of SC-GEM1, inducing an S2 EL signal. A large fraction of the resulting photons impinge on the photocathode of SC-GEM2; they extract photoelectrons that, in turn, generate another EL signal in the bubble underneath (S2'). A small fraction of the S2 photons emitted in the direction of the second electrodes (~18%) pass through the holes of SC-GEM2, reaching directly the bottom PMT. If a transfer field is applied between the two LHM elements, after generating the S2 signal, the ionization electrons drift to the SC-GEM2 element where they generate another EL signal (S3).

During the measurements it was noted that after an occasional intense discharge in our electrode, its maximal achievable voltage dropped significantly. As a result, the maximal voltage per element along this preliminary double-stage study was limited to ~750 V (it was 1300 V before the first few discharges). We note that in this phase of study we were driving the electrodes to the highest achievable voltage to investigate the safe limit of operation. However in normal operation, at voltages slightly below the limit, no violent discharges were observed.

## 3. Results

### 3.1 Horizontal LHM

#### 3.1.1 Waveform structure

As discussed above and shown schematically in Figure 3, for the CsI-coated LHM electrodes the waveform recorded by the bottom PMT, for each alpha particle emitted into the liquid, comprises three signals: (1) primary-scintillation light passing through the electrode holes (S1); (2) EL light generated inside the bubble by photoelectrons extracted from CsI following the absorption of primary scintillation photons in it (S1'), and (3) EL light generated inside the bubble by ionization electrons liberated along the alpha-particle track, which drift towards the LHM and are focused into its holes – inducing EL in the bubble (S2). Sample waveforms from the four electrodes are shown in Figure 6. The nominal drift field in Figure 6 was $E_d = 0.5$ kV/cm for all electrodes. The nominal transfer field (between the electrode and the resistive-wire plane) was $E_t = -1$ kV/cm, drifting the electrons towards the bottom face of the LHM-electrode after passing through the hole. The bottom PMT was operated at $-600$ V to avoid signal saturation at the Phillips 777 amplifier (limited at its input to 0.7 V due to protection diodes). At this voltage, the gain is ~10-fold lower than at the recommended operating voltage of -800 V, and therefore single electron spectra could not be obtained under these conditions; the change in PMT gain was corrected for by a separate measurement of the PMT response to S1 signals as a function of its voltage. For voltage values above 1,800 V across the 125 μm-thick SC-GEM and for the studies of EL across the transfer gap, despite low PMT voltage, the signals still saturated the electronics. This necessitated further reducing the PMT gain (HV) and normalizing the signals' amplitude.

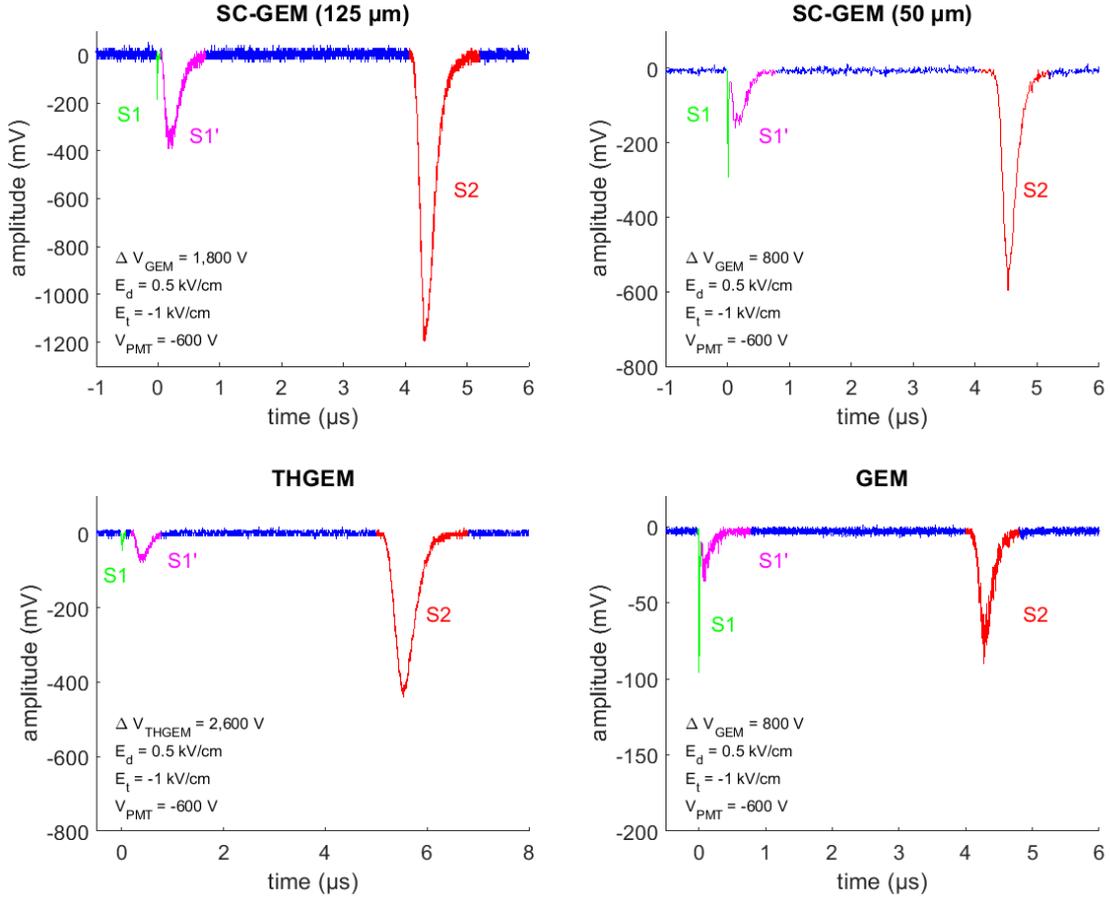

**Figure 6:** Sample single-event waveforms recorded from all four investigated electrodes by the bottom PMT, showing S1, S1' and S2. All waveforms were recorded at $E_d = 0.5$ kV/cm, $E_t = -1$ kV/cm and $V_{PMT} = -600$ V. The time difference between S1 and S1' (especially for the THGEM) can be used to infer that the bubble liquid-gas interface is located at the bottom of the hole (see Appendix A1).

The shape of the pulses bares information about the geometry of the structure. Specifically, the time difference between the onset of S1 and S1' equals the time it takes photoelectrons emitted from immediate vicinity of the hole to reach the bubble (liquid-gas) interface. Calculations of the drift lines originating at the photocathode and focused into the holes indicate that the liquid-gas interface is located at the bottom of the hole, with an apparently modest curvature as illustrated in Figure 2. Details of the analysis are given in appendix A1.

**3.1.2 Signal magnitude and energy resolution: comparison between different electrodes**

For given drift- and transfer fields the voltage across the electrode was varied gradually, with typically 10,000 waveforms recorded at each voltage step. The integrated pulse area under S2 and S1' was computed for each waveform, serving to build a histogram of pulse areas for each voltage setting. A Gaussian fit was applied to each histogram, giving the mean $\mu$ and standard deviation $\sigma$ of the S2 and S1' distributions. Histograms for the S1' and S2 signals of the 125 μm-thick SC-GEM, taken under the voltage configurations which yielded the best RMS resolutions, are shown in Figure 7. The Gaussian fit to the S1' distributions was applied over the range $\sim[\mu - \sigma, \mu + 4\sigma]$; it was done to exclude the low-amplitude contribution of events, which we attribute to partial energy deposition inside the source substrate by alpha-particles emitted at shallow angles (a similar low-energy tail was observed when measuring the source directly with a surface barrier detector, indicating that this effect does not result from LHM-related issues such as partial charge collection). When fitting the S2 data we also excluded the 'hump' on the right side of the peak, interpreted as coincident emissions of alpha particles and 59.5 keV gamma rays from $^{241}$Am. The resulting RMS values for S2 and S1' were 5.5% and 6.5%, respectively, for $\sim 7 \times 10^3$ primary electrons and $\sim 2 \times 10^3$ VUV-induced photoelectrons per event. The number of primary electrons was estimated from the energy of the alpha particles (5.48 MeV), and the number of electrons escaping recombination per keV from an alpha-particle track with $E_d = 0.5$ kV/cm (~1.3 electrons/keV [17]); the number of photoelectrons was estimated from the S1'/S2 ratio.

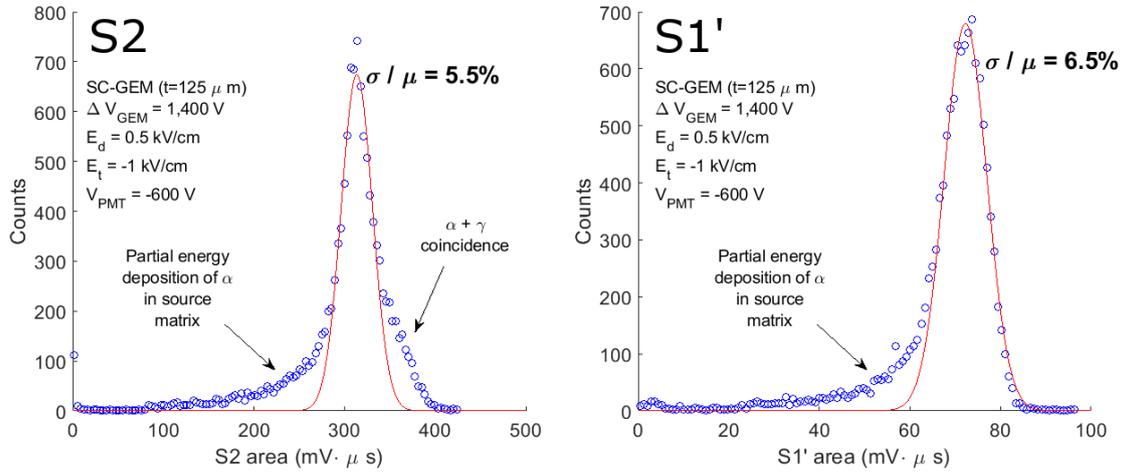

**Figure 7:** LHM EL spectra with a 125 µm-thick SC-GEM in LXe, of S2 induced by $\sim 7 \times 10^3$ primary electrons (left) and S1' induced by $\sim 2 \times 10^3$ photoelectrons (right). The excess of events to the left of the peak in both cases is attributed to partial energy deposition by alpha particles in the source substrate. The excess to the right of the S2 peak is due to coincident emission of alpha particles and 59.5 keV gamma rays. The Gaussian fit is applied over the range $\sim[\mu - \sigma, \mu + 4\sigma]$, excluding the low-energy tail and the excess of events right of the S2 peak.

Figure 8 shows the mean magnitude $\mu$ (integrated pulse area) of S2 and S1' as a function of the voltage across the four investigated electrodes, for the above nominal drift- and transfer fields. The data in Figure 8 are given in 'raw form' (i.e., measured pulse area). An estimate of the maximal effective EL yield (number of EL photons per electron entering the electrode holes, emitted over 4π), ~400 photons/e⁻/4π, is added for the highest point of the 125 µm-thick SC-GEM curve. The effective light yield was estimated by dividing the S2 pulse area by the product of the PMT's response to single photoelectrons, the number of primary electrons originating from the interaction, the probability that an EL photon reaches the PMT and the PMT's QE (see detailed description in Appendix A2). The discontinuity of the 125 µm-thick SC-GEM curve at 1,900 V results from saturation of the PMT (starting at ~1,700 V); data points above 1,800 V were taken with a lower PMT voltage ($-550\ V$), and re-scaled based on the measured PMT response to S1 signals, as explained in Appendix A2.

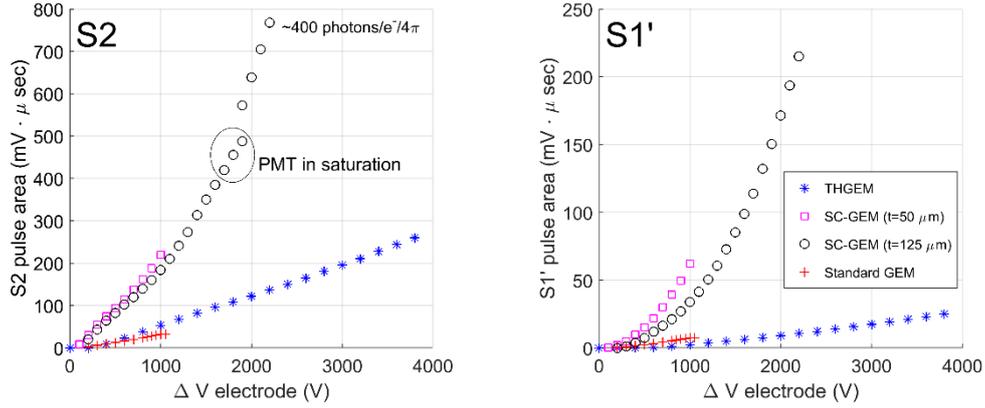

**Figure 8:** S2 (left) and S1' (right) pulse area measured as a function of the voltage set across the LHM electrode for the GEM, SC-GEMs and THGEM. In all configurations: $E_d = 0.5$ kV/cm and $E_t = -1$ kV/cm. The PMT was operated at $-600$ V (or $-550$ V for the 125 µm-thick SC-GEM at $\Delta V_{LHM} > 1,800$ V; signals were scaled accordingly for direct comparison). A rough estimate for the effective EL yield (number of photons per electron entering the hole over $4\pi$) is displayed at the maximal pulse-area point. Data taken at ~173 K and 1.6 bar.

At a given voltage, S2 magnitude is largest for the 50 and 125 µm-thick SC-GEMs. The latter, however, can be stably operated at much higher voltages, and its maximal S2 is ~3.5 larger than that of the 50 µm-thick SC-GEM, ~3 times larger than that of the THGEM and ~20 times larger than that of the standard GEM. We interpret the deviation from linearity starting at ~1 kV for the 125 µm-thick SC-GEM (not visible for the other electrodes) as mild avalanche multiplication within the bubble, with a charge gain of the order of ~2 at 2,200 V (In the absence of avalanche multiplication the EL yield depends linearly on the voltage, as verified by calculations done with COMSOL Multiphysics [19]; to estimate the gain at 2,200 V we take the ratio between the measured data point and the value expected by linear extrapolation from points corresponding to voltages below 1 kV.)

The measured S1' signal magnitude is also the largest for the 125 µm-thick SC-GEM: up to ~10 times larger than for the GEM and THGEM at the same applied voltage. The maximal value of the 50 µm SC-GEM S1' is ~2.5 times larger than that of the THGEM. For the S1' curves, the departure from linearity is attributed to two effects: small avalanche multiplication (similar to S2, but only for the 125 µm-thick SC-GEM) and an increase in the field on the surface of the CsI photocathode, leading to a higher overall photoelectron extraction efficiency into the liquid (similar to [18], in gas). The much larger measured value of S1' for the 125 µm-thick SC-GEM compared to the THGEM is attributed to the much higher surface field of the former (as verified by calculations done with COMSOL. Note that in light of the low alpha-particle emission rate (~90 Hz) and the very low (if any) avalanche multiplication, charging up effects during typical measurements lasting 1-2 hours should be minimal and can be neglected (e.g. see [20, 21]).

The ratio S1'/S2 of the THGEM, GEM and SC-GEMs is shown in Figure 9 as a function of the voltage across the four electrodes; as before $E_d = 0.5$ kV/cm and $E_t = -1$ kV/cm. This ratio should be, to zero order, proportional to the ratio between the numbers of S1' photoelectrons and S2 ionization electrons crossing the bubble interface without being lost to the electrode walls. The observation that this ratio is the largest for the two SC-GEMs, intermediate for the GEM and the lowest for the THGEM, suggests that the same order should hold for the photon detection efficiency (PDE) of the four types of CsI-coated electrodes. Direct measurements of the LHM PDE, using a dedicated setup, are underway and will be reported elsewhere.

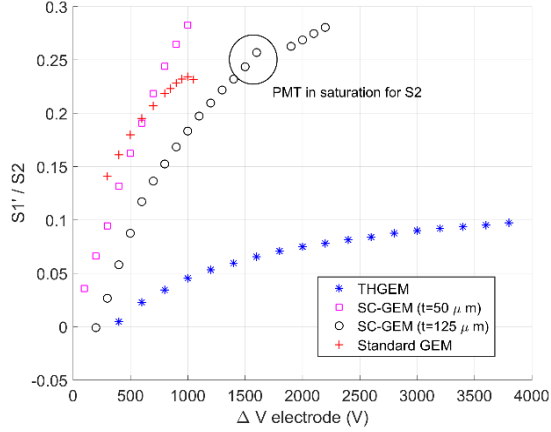

**Figure 9:** The ratio between the magnitudes (pulse areas) of S1' and S2 as a function of the voltage across the LHM electrode, for the GEM, two SC-GEMs and THGEM. $E_d = 0.5$ kV/cm, $E_t = -1$ kV/cm and $V_{PMT} = -600$ V (except for the 125 μm-thick SC-GEM, for which $V_{PMT} = -550$ V above 1,800 V). The encircled data points were recorded with the PMT partially saturated for S2 and therefore their deviation from the general trend is an artifact.

The RMS resolution $\sigma/\mu$ of S2 and S1' EL signals as a function of the voltage across the electrode is shown in Figure 10, for $E_d = 0.5$ kV/cm, $E_t = -1$ kV/cm and $V_{PMT} = -600$ V ($-550$ V for the 125 μm-thick SC-GEM, for $\Delta V_{LHM} > 1,800$ V). Initially the resolution improves as a function of the voltage, indicating a larger number of collected ionization electrons (S2, Figure 10 left) and photoelectrons (S1', Figure 10 right). The best S2 resolution is ~5.5% RMS for the 125 μm-thick SC-GEM.

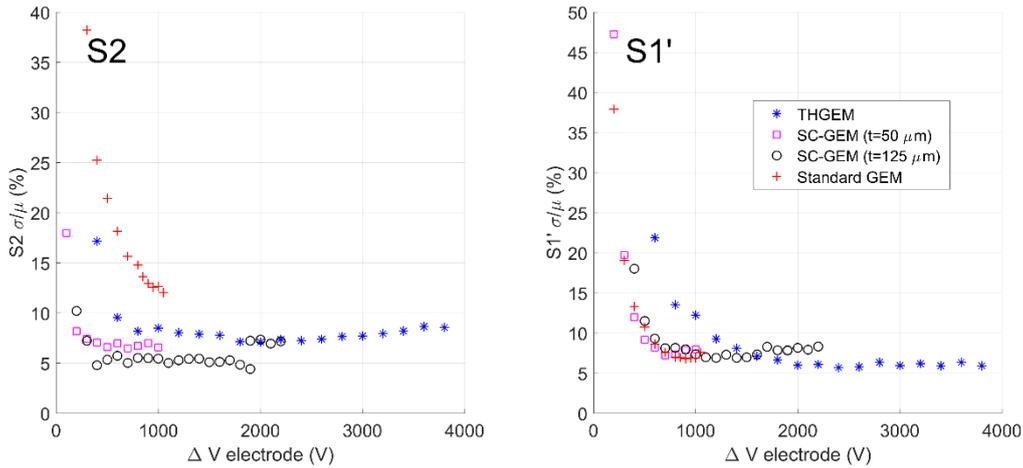

**Figure 10:** RMS resolution of the Alpha-particle induced pulse distributions of S2 (left) and S1' (right), in LHM configurations having THGEM, GEM and SC-GEM electrodes, as a function of the voltage across the electrode. $E_d = 0.5$ kV/cm, $E_t = -1$ kV/cm and $V_{PMT} = -600$ V ($-550$ V for the 125 μm-thick SC-GEM, for $\Delta V_{LHM} > 1,800$ V). The abrupt change of S2 resolution for the 125 μm-thick SC-GEM at 1,900 V results from the change of PMT voltage.

### 3.1.3 Drift field effect on S1' and S2

Figure 11 shows the effect of the drift field $E_d$ on the S2 (left) and S1' (right) magnitude, for all four electrodes. While the magnitude of S2 increases with $E_d$ as more ionization electrons escape recombination along the alpha-particle track [17] and make it to the electrode, that of S1' decreases with $E_d$ because the overall extraction efficiency across the SC-GEM top surface decreases (the net surface field decreases with $E_d$).

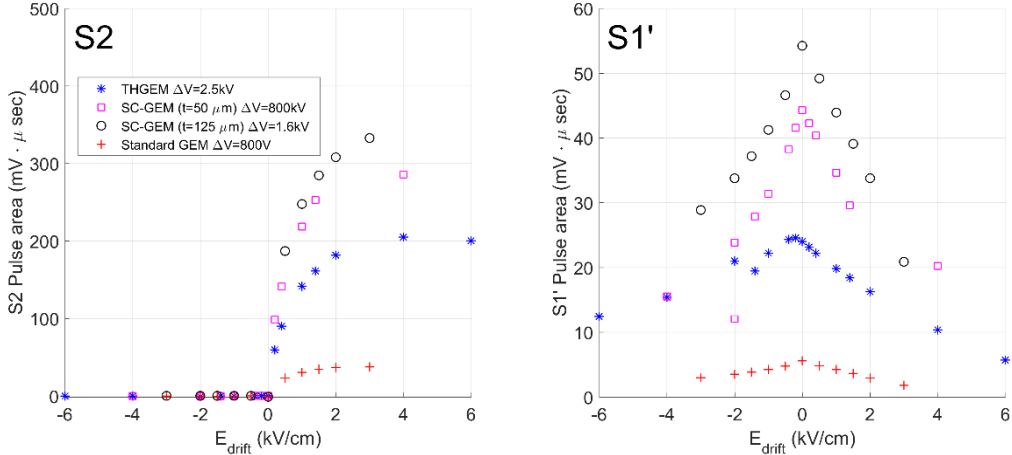

**Figure 11:** Effect of the drift field on S2 (left) and S1' (right) pulse area for the all four electrodes. Pulses recorded with $E_t = -1$ kV/cm and $V_{PMT} = -600$ V; voltages across the different electrodes are indicated in the legend.

### 3.1.4 Light amplification in the transfer gap

The threshold EL field in Xe vapor at 1.5 bar is 2.4 kV/cm [12]. At modest transfer fields between the bottom of the LHM electrode and wire plane (Figure 3), the only region in which the local field is larger than the EL threshold is inside the bubble, at the bottom of the hole (just below the liquid-gas interface). A major enhancement in the EL yield can be obtained by increasing the transfer field, pulling the electrons towards the wires. At first, additional EL occurs close to the wire surface, where the local field grows as $1/r$; at higher transfer voltages the field across the entire gap exceeds the threshold and EL light is generated along the full trajectory of the electrons as they drift across the bubble.

This effect is demonstrated in Figure 12 for the 125 μm-thick SC-GEM biased at 1,800 V with $E_d = 0.5$ kV/cm. The nominal transfer field $E_t$ was varied in steps between zero and 15 kV/cm. The bottom PMT was operated at -450 V to avoid signal saturation. Figure 12 shows the average waveforms acquired at increasing transfer field values. Initially, the waveforms look similar to those in Figure 6, comprising S1, S1' and S2 signals, where the latter two are produced in the bubble, close to the bottom of the hole. Increasing $E_t$, one observes delayed EL contributions appearing after both S1' and S2 ("S1'-echo" and "S2-echo"). These are EL signals produced inside the bubble, in the vicinity of the wires, in addition to those produced close to the holes. Further increase of the transfer field above the EL threshold along the transfer gap itself, leads to a merger between the EL signals produced near the holes and at the wires. The unified EL signals grow in magnitude and shrink in width, as $E_t$ is set to larger values (the width of the pulses decreases because of the increase of electron drift velocity in Xe gas [22]).

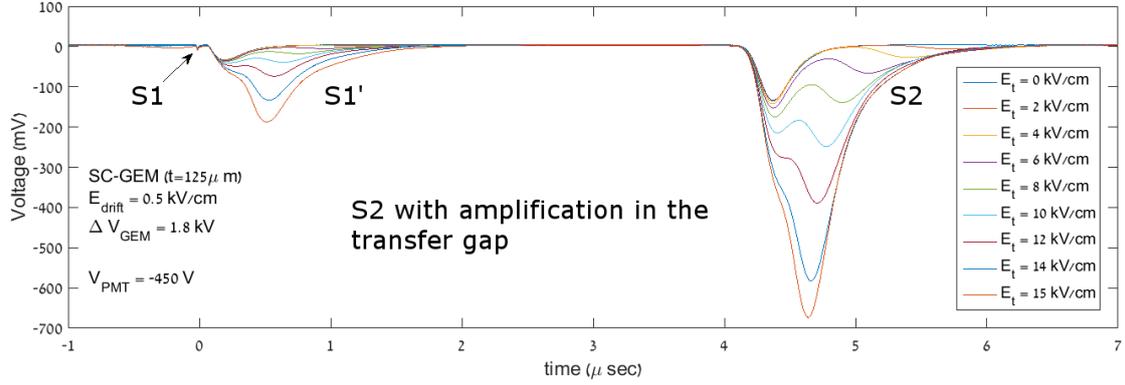

**Figure 12:** Average waveforms for increasing values of the nominal transfer field between the bottom of a 125 μm-thick SC-GEM operated at 1,800 V and the resistive-wire plane (located 1.6 mm below the GEM). $E_d = 0.5$ kV/cm and the bottom PMT voltage is $-450$ V to avoid signal saturation at large light signals.

The magnitudes of the amplified S1' and S2 signals, as a function of the transfer field, are shown in Figure 13 on a logarithmic scale. For S2 of the 125 μm-thick SC-GEM, the magnitude increases by 4,000 mV·μs over the scanned $E_t$ range, which is a factor ~8 larger than the S2 at 1,800V with zero transfer field. In terms of effective light yield, this translates to $\sim 2 \cdot 10^3$ photons per electron entering the LHM hole, over $4\pi$. The pronounced exponential trend of the S1' and S2 magnitude curves indicates a considerable avalanche charge gain on the wires. The data were recorded at different PMT voltages (from $-450$ V to $-600$ V) to accommodate both small and large pulses; the signals were scaled to be presented as if all were recorded at $-600$ V. The energy resolution here was worse than in the previous studies (~6-12%). However, this is likely an artifact as reduction in the PMT voltage is always followed by a change in its energy resolution, especially at such voltages which are far from its recommended operational voltage ($-800$ V). The effect is demonstrated best in Figure 10, where for the same voltage configuration (SC-GEM biased at 1,800 V), the resolution changes abruptly from ~5% to ~8% as the PMT voltage is changed from $-600$ V to $-550$ V.

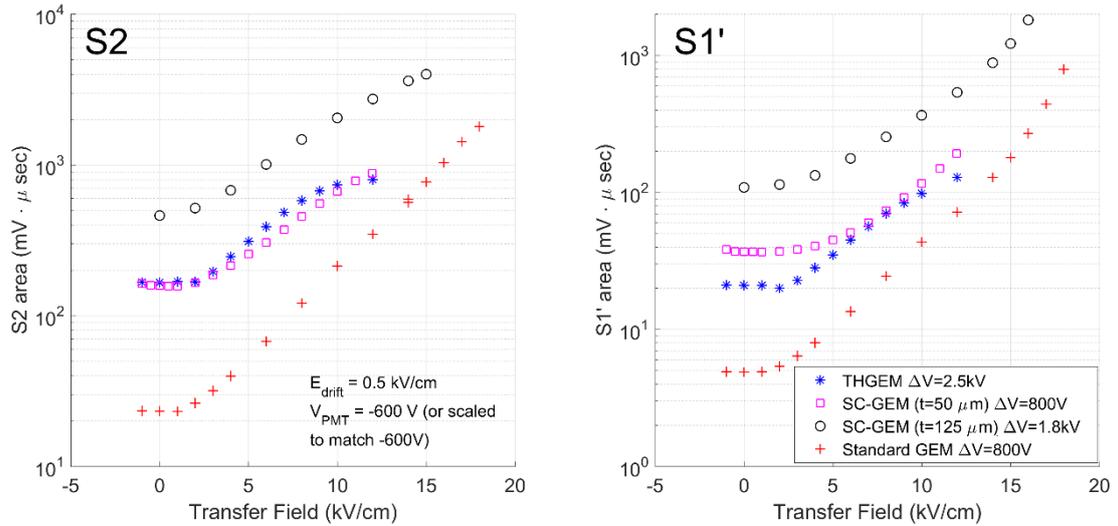

**Figure 13:** Magnitude of S2 and S1' of all four electrodes, as a function of the transfer field applied across the bubble between bottom face of the electrode and the wire plane. Signals recorded at $E_d = 0.5$ kV/cm.

Further information about electron transfer from the liquid phase into the bubble can be obtained from careful analysis of light generation in the transfer gap. At sufficiently large transfer fields, all of the S2 electrons which pass through the liquid-gas interface drift to the wire mesh (and not to the bottom face of the electrode). Under such conditions, the largest contribution to the S2 signal comes from the transfer gap and from the wires rather than from the bottom of the hole of the electrode. To zero-order approximation, light generation in the gap and on the wires should not depend on the specific details of the electrode and will be identical for all setups, provided that the same amount of charge is being transferred into the bubble. All setups in this study used the same alpha source, same drift field and - at the voltages applied to the electrodes - show full collection of electrons from the drift region into the hole (verified by COMSOL simulations). Therefore, the signal which is generated in the transfer gap becomes a measure of the relative transfer efficiency of electrons from the liquid into the bubble.

We now turn to carefully analyze the data. We choose to study the effect at $E_t = 8$ kV/cm, where COMSOL simulations show that all electrons crossing the liquid-gas interface reach the wires, and where the PMT is not saturated (Figure 13). The signal at zero transfer field is subtracted from the signal at 8 kV/cm to separate the contribution of the transfer gap from that of the electrode. Then, one should note that at 1,800 V across the 125 μm-thick SC-GEM, there is a small charge gain (estimated to be ~1.5 from the deviation from linearity presented in Figure 8 with possible enhancement by the superposition of the hole dipole field and the transfer field), for which the signal should be corrected in order to reconstruct the original measure for the charge. For comparison, we set the signal of the electrode with the highest pulse (the 125 μm-thick SC-GEM) to be 1 and compute the relative charge transfer efficiencies of all the other electrodes. The results are shown in Table 2.

**Table 2:** relative electron transfer efficiency from liquid to gas (normalized to that of the 125 μm-thick SC-GEM)

| Electrode | Normalized relative electron transfer efficiency |
|---|---|
| 125 μm-thick SC-GEM | 1 |
| 50 μm-thick SC-GEM | 0.44 |
| THGEM | 0.44 |
| Standard GEM | 0.14 |

This dramatic difference (of up to a factor of 7) in the transfer of electrons from above the electrode to the bubble is currently subject for further research and may explain, in part, the apparent difference between the light yields of the different electrodes. A possible hypothesis for the mechanism causing this difference is suggested in the discussion section.

## 3.2 Vertical LHM

We now turn to the first results obtained with a vertical bubble, confined between a standard-GEM and a fine mesh (Figure 4). As noted in Section 2.2, triggering was done on S2 signals which were amplified by an intense transfer field set across the bubble. Figure 14 shows a typical waveform recorded with $\Delta V_{GEM} = 900$ V, $E_d = 1$ kV/cm, $E_t = 15.5$ kV/cm and $V_{PMT} = -900$ V.

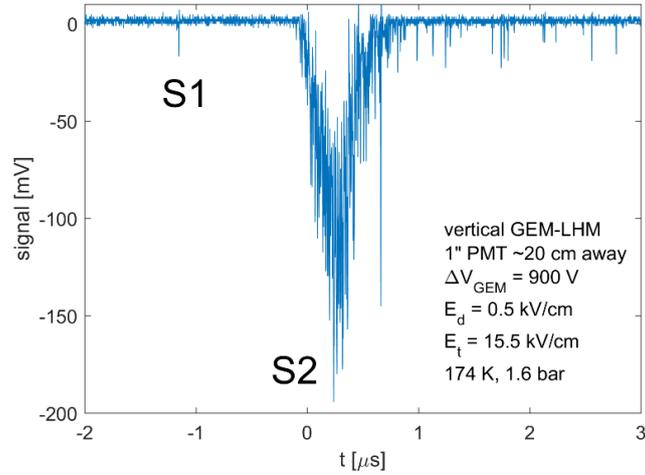

**Figure 14:** Single-event EL signal from the holes of a vertical bare standard-GEM electrode (without CsI) with a bubble sustained between the GEM and a fine-pitch mesh (see Figure 4). Photon statistics are low because of geometrical constraints which forced placing the 1" PMT outside of the vessel ~20 cm away from the LHM.

Figure 15 shows the average pulse area and RMS resolution of the S2 signals as a function of $E_t$; for each value of $E_t$, 10,000 waveforms were digitized and processed. At the highest light yield recorded under these non-optimal conditions, the resulting RMS resolution of 11% is worse than that achieved for the horizontal LHM, however this is probably due to the low light collection into the PMT at this specific setup.

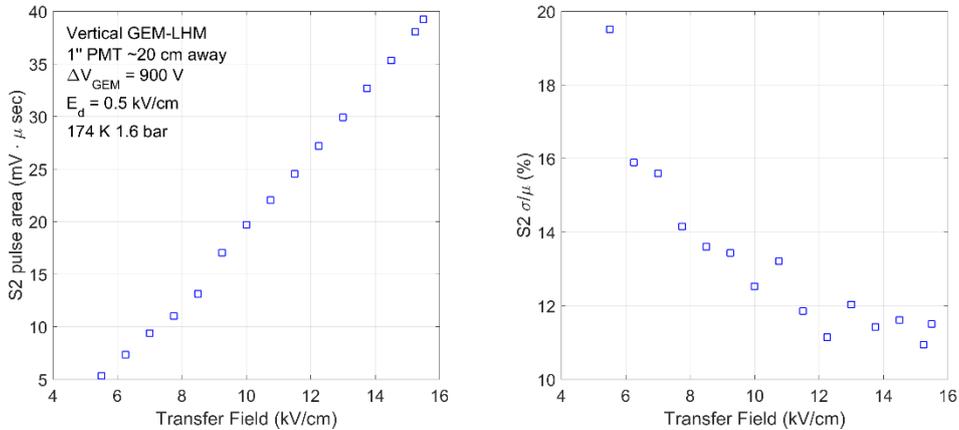

**Figure 15:** Vertical LHM with a standard-GEM. Relative S2 magnitude (A) and energy resolution (B) resulting from EL signals recorded from the vertical bubble confined between the GEM electrode and a fine mesh (setup of Figure 4). The EL photons originate from the high electric field applied across the transfer gap.

### 3.3 Double-stage LHM

In the first set of measurements of the double-stage SC-GEM LHM (300 μm pitch, 150 μm-diameter top hole, 50 μm thickness) (Figure 5), we studied the dependence of the light output of the structure on the voltage across the second electrode $\Delta V_{SC-GEM2}$, for a fixed voltage across the first stage $\Delta V_{SC-GEM1} = 700$ V, with $E_d = 1$ kV/cm and $E_{t1} = E_{t2} = 0$. Figure 16 shows an

– 16 –

average of 10,000 waveforms recorded by the bottom PMT at $V_{PMT} = -700$ V, for several values of $\Delta V_{SC-GEM_2}$. A small fraction of prompt scintillation photons (S1) traversing both electrodes reach the bottom PMT. The S1 signal is attenuated by light absorption by the electrodes (the relative area of the holes on the electrode top surface is 23%) and by total internal reflection from the liquid-gas interfaces (as discussed in [3]). The S1' signal in Figure 16 is due to photons that passed through the first electrode, extracting photoelectrons from the CsI on the second stage and inducing EL in the second bubble. The S2 signal resulting from electron-induced EL photons in the first bubble is followed by S2' EL, due to S2 photons extracting photoelectrons from the CsI; the latter induce further EL signals in the second bubble. Since S2 photons are emitted (in this geometry) during ~1 μs and since it takes ~30-50 ns [23] for a photoelectron in LXe to drift from the photocathode to the bubble of the second stage (across the 50 micron hole length), the S2 and S2' signals overlap. Thus, S2' is seen as an amplified (and slightly delayed) S2.

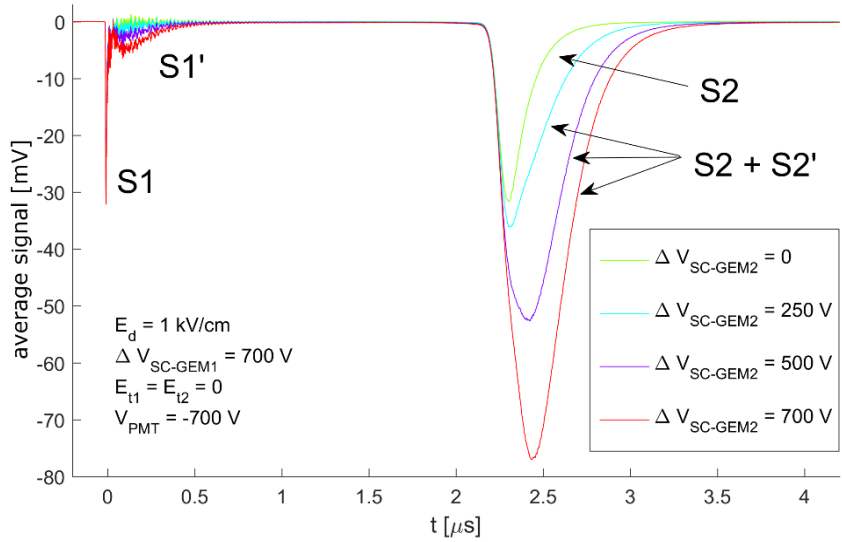

**Figure 16:** Averaged alpha-particle waveforms recorded by the bottom PMT in the double-stage, SC-GEM LHM setup (of Figure 5). Different colors correspond to different voltages applied across the second stage. EL S2 photons from the first stage induce the emission of further photoelectrons in the second stage. These generate additional EL S2' photons – thus amplifying the original signal.

For each waveform, the integral of the S2+S2' pulse was computed, deriving pulse-area histograms. An example (recorded at $E_d = 1$ kV/cm, $\Delta V_{SC-GEM1} = 700$ V, $\Delta V_{SC-GEM2} = 750$ V, $E_{t1} = E_{t2} = 0$ and $V_{PMT} = -700$ V) is shown in Figure 17. As in Figure 7 (left), the main peak is due to the 5.5 MeV alpha particles and the left tail is due to alpha particles leaving a fraction of their energy in the source substrate. The 'hump' on the right is due, at least in part, to alpha particles emitted in coincidence with 59.5 keV gamma photons; this effect, however, appears to be somewhat more pronounced here than for the single-stage LHM, calling for further investigations.

A Gaussian fit was applied to the S2+S2' spectra, for which the mean and the RMS resolution are shown in Figure 18 as a function of $\Delta V_{SC-GEM2}$. At high voltages the behavior is linear; the curved response at lower voltages is due to increasing photoelectron extraction efficiency from CsI with increasing electric field at the LHM surface (verified by electric-field maps calculated using COMSOL simulations and compared to the QE curve measured in [24]).

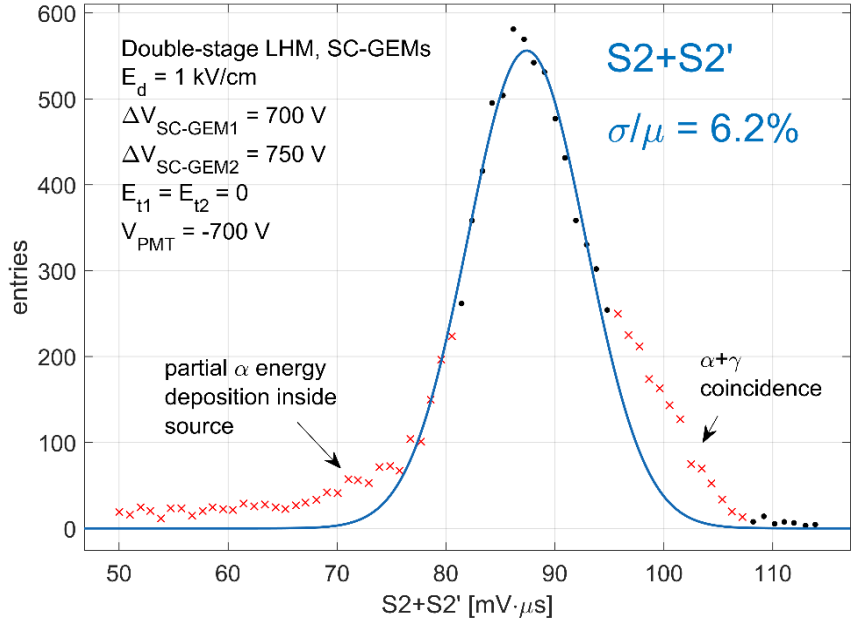

**Figure 17:** Double-stage SC-GEM LHM spectrum of S2+S2'. Data points in red were excluded from the Gaussian fit. The excess of events left of the peak is attributed to partial energy deposition by the alpha particle inside the source substrate, and that right of the S2 peak - to coincident emission of alpha particles and 59.5 keV gammas. The estimated number of ionization electrons entering the first LHM electrode is ~10,000.

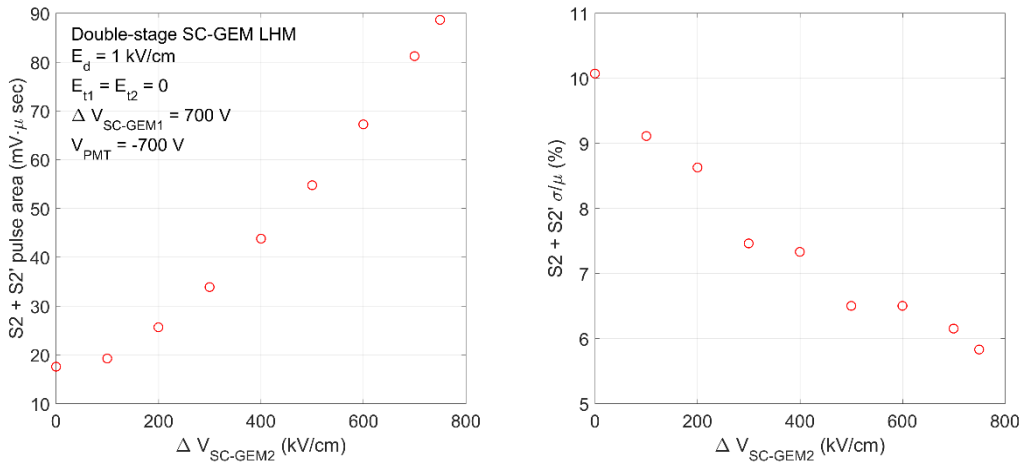

**Figure 18:** Double-stage SC-GEM LHM: S2 + S2' pulse area and RMS resolution as a function of the voltage across the second amplification stage.

We define the effective gain of the second stage to be the ratio between the combined S2+S2' pulse area and that of the first stage only (i.e., of a single-stage LHM configuration). In the current setup, the pulse-area of the first stage, when seen by the bottom PMT, is attenuated twice: the optical transparency of the second SC-GEM accounts for ~77% opacity (estimated from the electrode geometry) and total internal reflection at the liquid-gas interface accounts for further

~10% (the pulse-area of S1 without the second bubble, detected by the bottom-PMT, was compared to the S1 area with the second bubble). Considering the pulse area of 17.7 mV · μs at $\Delta V_{SC-GEM2} = 0$, an effective light transmission of ~20% through the holes of stage 2 (including reflections on the liquid-gas interface) and a pulse area of 88 mV · μs at $\Delta V_{SC-GEM2} = 750$ V, the effective gain of stage 2 was estimated to be ~1. When extrapolating the results of Figure 18 (left) to $\Delta V_{SC-GEM2} = 1400$ V (the maximal voltage obtained with these electrodes prior to discharges), one could potentially reach gains of ~2. This gain in light yield is much smaller than that obtained by applying an intense transfer field (gain of ~8, Section 3.1.4). However, the S2 RMS resolution of ~6% seems to remain similar to the resolution of a single element with low and inverted transfer field. It is worth mentioning that maintaining the signal resolution of an optically-coupled two-stage gas-avalanche detector was also observed previously [14].

Under a transfer field applied between both stages, electrons drift from the holes of the first electrode to that of the second one, where they induce an additional S3 EL signal. Figure 19 shows an average of 10,000 waveforms recorded at different voltage values of the first stage. In addition to the S2+S2' signals which obviously increase in amplitude with the voltage applied to the first electrode, the S3 signal is clearly apparent and shows a general trend of increase with $\Delta V_{SC-GEM1}$, likely indicating a minor increase in electron collection and/or transferring efficiency into the bubble of the first stage. Unfortunately, the data were taken after the electrodes have suffered from a few violent discharges, which resulted in instabilities in the detector's response. Therefore, consistent quantitative results could not be extracted from the current set of data.

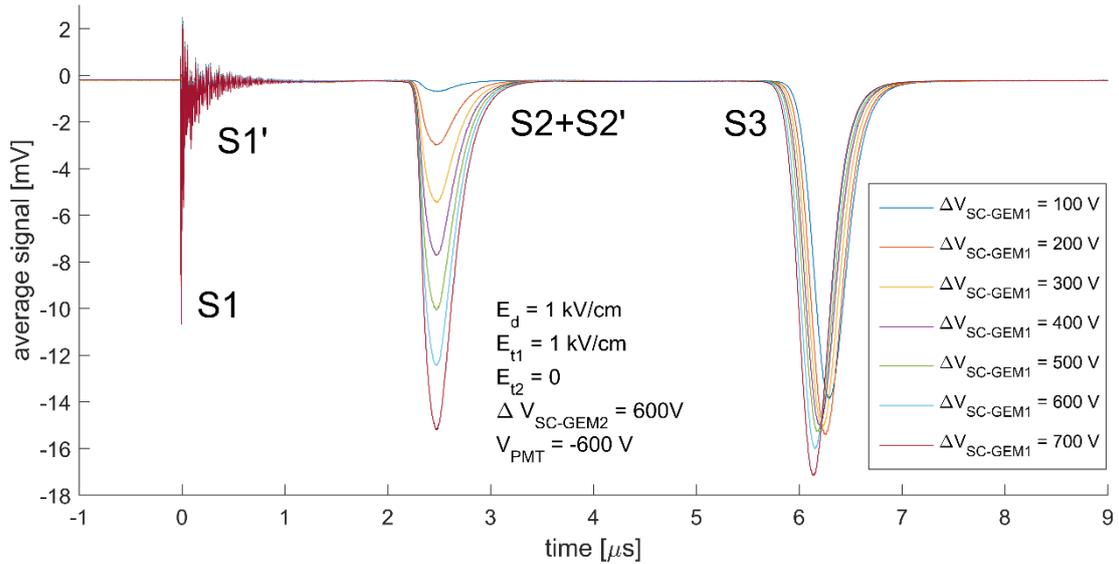

**Figure 19**: Average waveforms of pulses measured by the bottom PMT from the double-stage LHM (of Figure 5), with electrons drifting all the way to the second SC-GEM (under $E_{t1} = 1$ kV/cm). The S1 and S1' scintillation signals are followed by ionization-electron-induced ones, giving rise to the S2+S2' signal; S3 signals, due to electrons reaching stage 2, show a small increase in the pulse-area with $\Delta V_{SC-GEM1}$; it possibly originates from an increase in electron collection and/or transfer efficiency into the bubble of the first LHM stage.

## 4. Summary and discussion

In this work, we first established a systematic comparative study between four different LHM electrodes: a THGEM, a standard bi-conical GEM and two different single-conical GEMs (SC-GEMs). It was demonstrated that the electrode geometry plays a crucial role in determining the effective EL yield (number of EL photons emitted over $4\pi$ per electron entering the electrode's hole), and the resulting energy resolution. Currently, the 125 µm-thick SC-GEM (with 300 µm diameter upper holes) yielded the best results. Compared to the others, it shows more than 3-fold larger EL yield for both ionization electrons, and VUV photons.

The difference in the LHM-electrode's response to electrons is still a subject of extensive studies. One of the hypotheses recently raised focuses on the transfer of electrons through the liquid-to-gas interface. It is well known that electrons in noble liquid form a cloud of polarized molecules around them, leading to an effective potential well [12]. In LXe, the potential well depth is of 0.69 eV [12]. Thus, to pass from liquid to gas, an electron needs either to tunnel through this barrier or to be at the high-energy tail of a thermal distribution. In a parallel-plate configuration, such as in a classical dual-phase TPC, electrons are drifted under high electric field to the interface. Throughout this process, they are almost completely thermalized. Once they reach the interface, they remain there until they tunnel through the potential barrier under the intense electric field. In our case, electrons reach the liquid-to-bubble interface, encountering a similar potential barrier (liquid to vapor); an intense electric field "pushes" them against the interface, like in the dual-phase TPC. However, the field has a component tangent to the liquid-gas interface. In a semi-classical picture, the latter may cause the electrons to "glide" over the interface - causing them to either tunnel through the barrier or reach and be collected at the bottom of the hole-electrode. Electrons reaching the electrode bottom are lost and do not induce EL photons. The process is depicted in Figure 20. This hypothesis is under ongoing investigations.

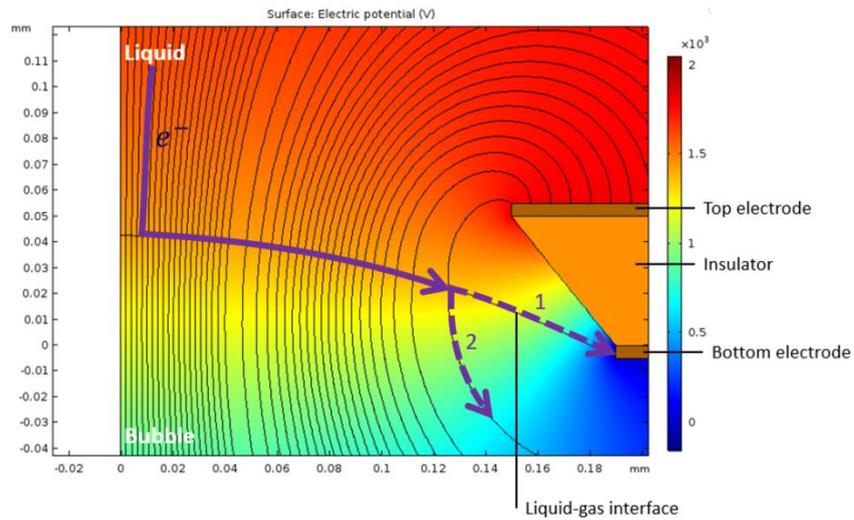

**Figure 20:** Simulation of electric field lines in the hole with a protruding spherical bubble. (1) is the proposed trajectory of electrons originating in the liquid, gliding on the liquid-gas interface and reaching the bottom face of the GEM. (2) is the proposed trajectory of electrons from the liquid, gliding on the interface and tunneling into the vapor phase.

The photon detection efficiency (PDE) has not been measured directly in the present work; it is however expected that SC-GEM electrodes will exhibit the highest PDE value among the electrodes investigated. This is based on their combination of a high surface field and relatively large holes – leading to a high photoelectron extraction efficiency from CsI into LXe [18] and reduced loss of electrons. Systematic studies aiming at direct measurement of the PDE are underway.

The RMS energy resolution obtained in the present work with the 125 μm-thick, 35 mm-diameter, SC-GEM is the best achieved so far in the context of the LHM studies, reaching 5.5% for ~7,000 deposited (S2) ionization electrons, and 6.5% for an estimated number of ~2000 VUV-induced (S1') photoelectrons (Figure 7). For comparison, the XENON100 collaboration reported 11%-13% S2 resolution for the same amount of charge, in a ~30 cm-diameter detector [25]. We attribute this, although further proof is necessary, to the tight confinement of the liquid-gas interface in the electrode hole. In a classical dual-phase TPC, S2 resolution may be sensitive to sagging and electrostatic deformation of the meshes, to surface waves and to instabilities of the liquid level. All these may combine to degrade the S2 resolution. In the case of the LHM, where the liquid-gas in confined in a single hole, all these effects are alleviated.

Detailed analysis of the data further helped in advancing our understanding regarding the location of the bubble interface at the bottom of the electrode. The time difference between the S1 and S1' signals, as well as the consistency between EL simulations and measurements, indicate that the liquid-gas interface is moderately curved, penetrating into the bottom of the hole.

A significant milestone is the demonstrated ability to enhance the EL yield to up to $\sim 2 \cdot 10^3$ photons over $4\pi$ per electron entering a hole – roughly 5-10 fold higher than conventional dual-phase TPCs [12, 25], albeit with some compromise of the energy resolution; it was reached by applying a ~15 kV/cm transfer field across a 1.6 mm thick bubble located underneath the electrode. This should lead to effective single-electron sensitivity in future LHM-detector modules. For example, if a 100 $cm^2$ LHM module is read out by an array of VUV-sensitive SiPMs of overall 5-10% detection efficiency (including dead spaces between the sensors), the number of SiPM-detected photoelectrons per single electron inducing EL signal will be ~50-100 within a time window of ~1 μs (the duration of the EL pulse across the transfer gap); the total number of SiPM dark counts in this time window will be below ~0.01 (assuming a typical dark-count rate of 1 Hz/$mm^2$ at LXe temperature [26, 27]). Furthermore, the detection of ~50-100 PEs on a pixelated SiPM array will allow for precise mm-scale localization of the event across the LHM active area; it will enable accurate position reconstruction and fiducialization in a TPC employing such modules, as well as the use of a position-correction map for energy (charge) measurements.

The study of the vertical GEM-LHM configuration is in its infancy. For now, it is yet to be demonstrated that a vertical bubble can be maintained stable in steady state. Still, the RMS energy resolution (~11%) obtained under the non-optimized conditions of the present experiments (namely, low photon collection efficiency), is surprisingly good.

While the results on the double-stage configuration are very preliminary, they provide a first demonstration that a cascaded structure with independent bubbles is indeed feasible. Although the present double-stage configuration was apparently limited to an overall EL amplification of ~2 - considerably lower than possible with a single-stage LHM with EL amplification in the transfer gap - it does show a very good energy resolution. Furthermore, the overall EL yield can in principle be enhanced by the addition of one or more stages. A potentially useful feature of the cascaded structure is the possibility of gating its response – for example by momentarily reversing the transfer field between the first and second electrode. This can add a new degree of freedom to LHM structures, enabling selective switching for detecting desirable events and being 'blind' to others.

Having demonstrated the bubble-assisted LHM as a potential solution for a combined detector of ionization electrons and scintillation photons, one could imagine how this may affect the design of future large-volume detectors. As discussed in [9] and depicted in Figure 21, one may consider a single-phase TPC configuration in which CsI-coated, position sensitive LHM modules are tiled at the bottom (with ionization electrons drifting downward). The main advantage of this configuration is that it can provide a highly uniform S2 response across the detector by maintaining the liquid-gas interface at a precise location below the LHM electrodes, together with an accurate position reconstruction capability. A further advantage is that, unlike conventional dual-phase noble-liquid TPCs, the liquid-gas interface is located outside of the sensitive volume. This reduces the amount of multiple reflections of S1 photons (by total internal reflection from the interface), thus improving S1 statistics at low energy depositions. A further improvement in S1 statistics is obtained by reducing the number of grids (which partially absorb photons) from five in the conventional dual-phase TPC scheme to two in the single-phase LHM-TPC design. Obviously, to be relevant for large-volume TPCs the LHM modules must be much larger in size compared to the present study (e.g., 10-20 cm diameter). Bubble stability, critical for the successful operation of the LHM, must therefore be demonstrated for such large modules, and under realistic operation conditions (in particular, under high hydrostatic pressure). This will be a central subject of near-future investigations.

Although the original motivation for the LHM idea was an attempt to solve difficulties associated with the scaling up of noble-liquid dark-matter detectors to the multi-ton regime, this concept may potentially find applications in other fields as well. For example, one may consider applying this idea in large LXe TPCs for neutrinoless double beta decay searches. Furthermore, although the LHM concept has so far only been demonstrated in LXe, we expect it to hold also in LAr, which may potentially open further applications in neutrino physics experiments. One may further consider a bi-directional dual-phase noble-liquid TPC with a central horizontal cathode, charge readout on top (in the vapor phase, or on immersed wires) and combined charge and light readout using LHM modules at the TPC bottom. Such a design has the potential merit of reducing the drift distance and applied voltage (and hence required electron lifetime) by a factor of two compared to current designs. Other potential applications can be for small-scale noble-liquid TPCs as neutron and gamma detectors for low-rate nuclear physics experiments, e.g., with rare isotope beams. Such TPCs can offer high detection efficiency, excellent position reconstruction, good (ns-scale) timing capability and excellent neutron-gamma discrimination.

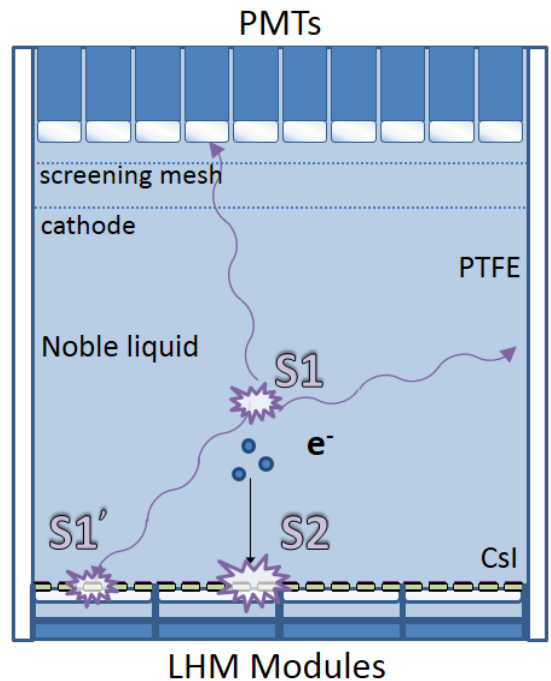

**Figure 21:** Conceptual scheme of a large-scale single-phase noble-liquid TPC employing CsI-coated bubble-assisted LHM modules, as in Figure 1. The LHM modules constitute the anode plane and are sensitive to both the ionization electrons (which, in this scheme, drift downward) and S1 scintillation photons. There is no need for additional grids between the LHM modules and the TPC drift volume; only two grids are required – the cathode and a screening mesh protecting the top array of photon detectors (e.g., PMTs). Such a scheme can potentially have a more uniform S2 response than 'conventional' dual-phase TPCs, as well as an improved light collection ('light yield') efficiency, due to the reduced number of reflections (with the liquid-gas interface outside of the sensitive volume) and due to the reduced number of photon-absorbing mesh electrodes (two instead of five).

## Appendix A: Further analysis

### A.1. Location of the bubble interface

The time delay between S1 and S1' for the single-stage LHM (Figure 6) can be used as a probe to estimate the location of the bubble top interface inside the electrode hole. The simplest case is that of the THGEM-LHM. Here, S1' lags after S1 by ~200 ns. This delay can be interpreted as follows. The starting time of S1' corresponds to the arrival of the first photoelectrons emitted from the CsI photocathode to the bubble interface. The shortest photoelectron trajectories are those starting from the CsI-coated rim (see Figure A1). This figure shows the electrostatic field calculated using COMSOL for a representative voltage configuration, assuming – as noted above (and shown in Figure 2) – that the bubble interface is spherical and located at the bottom of the hole. The field map is calculated on the plane crossing the THGEM-LHM unit cell along its diagonal (through the hole centers) and is overlaid by selected drift lines of photoelectrons emitted from the CsI-coated surface on the THGEM top face (as well as of ionization electrons originating from the drift gap). The ratio between the height of the bubble apex (relative to the bottom of the FR4) and the hole diameter was arbitrarily taken as 1:4. As discussed below, the particular choice of value for this ratio has little effect on the final conclusion.

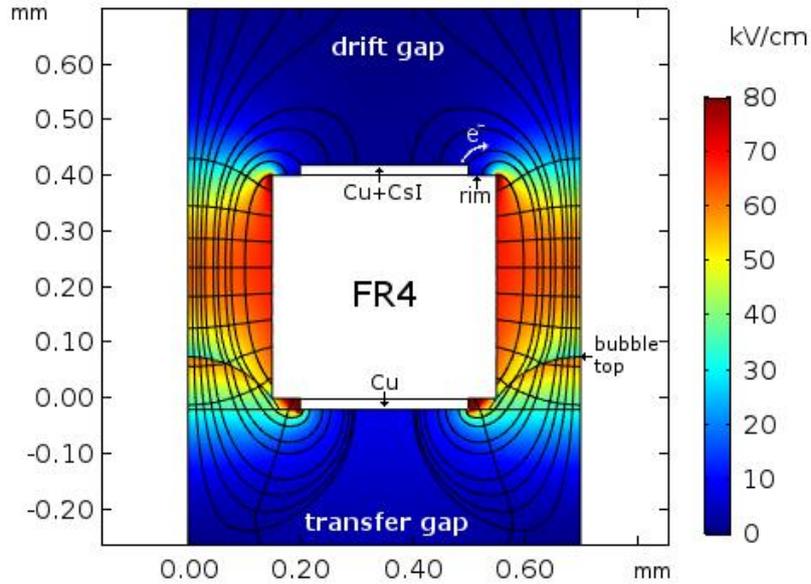

**Figure A1**: Electric field map across a plane crossing the THGEM unit cell (Figure 2) in diagonal, passing through the centers of the holes at opposite corners. The central white region is the electrode's FR4 substrate, with cuts through the holes on its sides. Spherical bubble interfaces are shown at the bottom of both holes. Also shown are field lines, describing ideal electron drift trajectories (with zero diffusion) starting either in the drift gap or on the top (CsI-coated) surface of the electrode. The electron trajectories end on the bottom conductive surface of the electrode due to the application of a reverse transfer field. In this particular case the voltage across the THGEM is 2900 V, the drift field is 0.5 kV/cm and the reverse transfer field is -2.8 kV/cm. In regions where the local field is larger than ~10 kV/cm, the drift time of a photoelectron along a particular trajectory, from the CsI surface to the bubble interface, is the ratio between the length of the trajectory and the essentially constant drift velocity.

The calculated length of the shortest photoelectron trajectory, starting on the inner edge of the rim (i.e., the edge of the hole) and ending on the bubble interface is 0.46 mm. (In fact, as can be seen in Figure A1, such a trajectory ends on the bottom part of the FR4 wall, if no charging-up effects are assumed). Since the field is larger than 10 kV/cm at all points along this trajectory, and the electron drift velocity in LXe saturates at 2.8 mm/μs at such fields [23], the corresponding drift time is $t_d \approx 0.46/2.8\,\mu s \approx 160$ ns. Similarly, the corresponding drift time for an electron starting at the outer edge of the rim (from the edge of the conductive surface) is ~220 ns, and for an electron starting at the mid-point of the rim ~210 ns. The calculated drift times are fairly insensitive to the assumed curvature of the bubble interface, as long as the height/diameter ratio is not too large. However, if one assumes that the bubble penetrates deep into the hole, the calculated drift times become much shorter and do not agree with the measured S1-S1' delay. It therefore appears, that at least for the THGEM, the bubble interface is indeed close to the bottom of the hole. A similar argument can be made for the GEM and SC-GEMs, where the shortest trajectories from the edge of the top Cu surface to the bottom of the hole correspond to a drift time of ~20-30 ns; in this case, however, the time lag between S1 and S1' is less clear, because S1' begins during the exponential decay of S1 and the signals cannot be easily separated.

## A.2. Absolute EL yield

The effective EL yield of the LHM electrodes, i.e., the number of EL photons emitted per electron focused into the electrode hole, over the full solid angle, can be estimated as follows. We denote by $N_e(E_d)$ the number of ionization electrons escaping recombination along the alpha-particle track, which depends on the drift field $E_d$; $\varepsilon(\Delta V_{LHM}, E_d, E_t)$ is the overall efficiency of focusing electrons into the hole and subsequently transferring them across the liquid-gas interface into the bubble, which depends primarily on the voltage across the electrode $\Delta V_{LHM}$ and on the drift field, and possibly also on the transfer field; $Y_{EL}(\Delta V_{LHM})$ is the number of photons emitted over $4\pi$ per electron *which enters the bubble*, and is assumed to depend primarily on $\Delta V_{LHM}$ (the effect of $E_d$ and $E_t$ on $Y_{EL}$ is assumed to be small); $P_{LHM \rightarrow PMT}$ is the probability that an EL photon emitted from the bubble reaches the PMT; $QE_{PMT}$ is the quantum efficiency of the PMT at 171 nm, i.e. the average number of photoelectrons (PEs) per photon hitting the PMT window. The number of PEs recorded by the PMT (i.e., emitted from its photocathode and contributing to the anode signal) is given by:

$$N_{PE} = N_e \cdot \varepsilon \cdot Y_{EL} \cdot P_{LHM \rightarrow PMT} \cdot QE_{PMT} \tag{A1}$$

We define the effective EL yield as the average number of EL photons emitted per electron escaping recombination at the alpha particle track:

$$Y_{EL}^{eff} = Y_{EL} \cdot \varepsilon = \frac{N_{PE}}{N_e \cdot P_{LHM \rightarrow PMT} \cdot QE_{PMT}} \tag{A2}$$

The number of PEs in a given EL signal is the pulse area of the signal divided by the average pulse area of a single-PE event, obtained from the single-PE spectrum of the PMT measured at LXe temperature. The average pulse area corresponding to the single-PE peak was measured at $V_{PMT} = -800$ V to be $0.03$ mV·μs. Since the data were typically acquired at $V_{PMT} = -600$ V, we further calibrated the PMT response at the different voltages. This was done by recording the S1 peak (in the setup shown in Figure 3, with no bubble present and zero voltages on all electrodes) for different values of $V_{PMT}$. The PMT S1 response at $-600$ V was found to be 0.1 that at $-800$ V, thus estimating $0.003$ mV·μs per photoelectron at $-600$ V.

The QE of the bottom-PMT was measured using a vacuum UV monochromator, versus a calibrated (NIST-traceable) photodiode. It was found to be 31% at 171 nm for normal incidence. Since this wavelength is close to the cutoff of fused silica, we expect the QE for photons incident at other angles to be somewhat smaller. The probability that a photon emitted from the bottom of the LHM reaches the PMT was calculated using a dedicated Monte-Carlo simulation (which included refraction and reflection on the bottom interface of the bubble), giving $P_{LHM \rightarrow PMT} = 0.32$.

We used the measured S2 signals in a single-LHM horizontal mode (Figure 8) as a basis for calculating the effective EL photon yield. Adopting the data from [17], for $E_d = 0.5$ kV/cm, the charge yield of $^{241}$Am 5.48 MeV alpha particles is $N_e = 7.1 \times 10^3$ electrons. Using equation A2 and the data of the 125 μm-thick SC-GEM, we find that the effective light yield at $\Delta V_{LHM} = 2,200$ V is ~400 photons/e⁻ over $4\pi$.

As discussed in Section 3.1.4 the effective EL photon yield can be boosted by applying an intense transfer field across the bubble. In particular, for the 125 μm-thick SC-GEM operated at at $\Delta V_{LHM} = 1,800$ V, and a nominal transfer field of 15 kV/cm applied over a 1.6 mm gap with 50 μm diameter wires, the overall effective light yield was 5-fold larger than obtained at 2,200 V with no contribution from the transfer field, i.e. ~$2 \times 10^3$ photons/e⁻ over $4\pi$.

## Acknowledgments

This work was partly supported by the Israel Science Foundation (Grant No. 791/15) and Minerva Foundation with funding from the German Ministry for Education and Research (Grant No. 712025). We are indebted to Dr. M. Rappaport of the Physics Core Facilities for his thoughtful advice and assistance on the cryogenic design of the experimental setup. We thank B. Pasmantirer and O. Diner from the Design Office, H. Takeia and members of his Mechanical Workshop and Y. Asher from the Physics Core Facilities (all at the Weizmann Institute) for their invaluable assistance in the design and manufacture of the experimental MiniX setup. We also thank Dr. A. Kish of Zurich University for the manufacture of the PMT bases and Dr. O. Aviv, L. Broshi, Z. Yungrais and T. Riemer from Soreq NRC for the preparation of the $^{241}$Am source. The research was carried out within the R&D program of the DARWIN Consortium for future LXe dark matter observatory. It is part of the CERN-RD51 detector R&D program.

## References

[1] A. Breskin, *Liquid Hole-Multipliers: A potential concept for large single-phase noble-liquid TPCs of rare events*, *J. Phys. Conf. Ser.* **460**(2013) 012020.

[2] L. Arazi, A. E. C. Coimbra, R. Itay, H. Landsman, L. Levinson, B. Pasmantirer, M. L. Rappaport, D. Vartsky and A. Breskin, *First observation of liquid-xenon proportional electroluminescence in THGEM holes*, 2013 *JINST* **8** C12004.

[3] L. Arazi, E. Erdal, A. E. C. Coimbra, M. L. Rappaport, D. Vartsky, V. Chepel and A. Breskin, *Liquid Hole Multipliers: bubble-assisted electroluminescence in liquid xenon.* 2015 *JINST* **10** P08015.

[4] E. Erdal, L. Arazi, V. Chepel, M. L. Rappaport, D. Vartsky and A. Breskin, *Direct observation of bubble-assisted electroluminescence in liquid xenon*, 2015 *JINST* **10** P11002.

[5] E. Erdal, L. Arazi, M. L. Rappaport, S. Shchemelinin, D. Vartsky, A. Breskin, *First demonstration of VUV-photon detection in liquid xenon with THGEM and GEM-based Liquid Hole Multipliers*, *Nucl. Instrum. Meth.* A **845**(2017) 218.

[6] F. Sauli, GEM: *A new concept for electron amplification in gas detectors*, *Nucl. Instrum. Meth.* A **386**(1997) 531.

[7] A. Breskin, R. Alon, M. Cortesi, R. Chechik, J. Miyamoto, V. Dangendorf, J. Maia and J. M. F. Dos Santos, *A concise review on THGEM detectors*, *Nucl. Instrum. Meth.* A **598**(2009) 107.

[8] M. Schumann, L. Baudis, L. Bütikofer, A. Kish and M. Selvi, *Dark matter sensitivity of multi-ton liquid xenon detectors*, *JCAP* **1510** (2015) 016.

[9] The DARWIN collaboration, *DARWIN: towards the ultimate dark matter detector*, *JCAP* **11** (2016) 017.

[10] T. Takahashi, S. Himi, M. Suzuki, J. Ruan (Gen) and S. Kubota, *Emission spectra from Ar-Xe, Ar-Kr, Ar-N2, Ar-CH4, Ar-CO2 and Xe-N2 gas scintillation proportional counters*, *Nucl. Instrum. Meth. 205 3 (1983) 591*.

[11] S. Duarte Pinto, M. Villa, M. Alfonsi, I. Brock, G. Croci, E. David, R. de Oliveira, L. Ropelewski and M. van Stenis, *Progress on Large-Area GEMs*, 2009 *JINST* **4** P12009.

[12] V. Chepel and H. Araujo, *Liquid noble gas detectors for low energy particle physics*, 2013 *JINST* **8** R04001.

[13] A. Breskin, *CsI UV photocathodes: history and mystery*, *Nucl. Instrum. Meth.* A 371 (1996) 116.

[14] J. F. C. A. Veloso, F. D. Amaro, J. M. F. dos Santos, A. Breskin and R. Chechik, *The Photon Assisted Cascaded Electron Multiplier: a Concept for Avalanche-Ion Blocking*, 2006 *JINST* **1** P08003.

[15] A. Buzulutskov and A. Bondar, *Electric and Photoelectric Gates for ion backflow suppression in multi-GEM structures*, 2006 *JINST* **1** P08006.

[16] A. Breskin, G. Charpak, S. Majewski, G. Melchart, A. Peisert, F. Sauli, F. Mathy and G. Petersen. *High flux operation of the gated multistep avalanche chamber*, *Nucl. Instrum. Meth.* **178** 11 (1980).

[17] E. Aprile, R. Mukherjee and M. Suzuki, *Ionization of liquid xenon by $^{241}$Am and $^{210}$Po alpha particles*, *Nucl. Instrum. Meth.* A **307** (1991) 119.

[18] L. Arazi, A. E. C. Coimbra, E. Erdal, I. Israelashvili, M. L. Rappaport, S. Shchemelinin, D. Vartsky, J. M. F. dos Santos, and A. Breskin, *First results of a large-area cryogenic gaseous photomultiplier coupled to a dual-phase liquid xenon TPC*, 2015 *JINST* **10** P10020.

[19] COMSOL Multiphysics® https://www.comsol.com/comsol-multiphysics, accessed September (2017).

[20] D. Shaked Renous, A. Roy, A. Breskin and S. Bressler, *Gain stabilization in Micro Pattern Gaseous Detectors: methodology and results*, 2017 *JINST* **12** P09036.

[21] M. Pitt, P. M. M. Correia, S. Bressler, A. E. C. Coimbra, D. Shaked Renous, C. D. R. Azevedo, J. F. C. A. Veloso and A. Breskin, *Measurements of charging-up processes in THGEM-based particle detectors*, 2018 *JINST* **13** P03009.

[22] H. L. Brooks, M. C. Cornell, J. Fletcher, I. M. Littlewood and K. J. Nygaard, *Electron drift velocities in xenon*, *J. Phys. D: Appl. Phys.* **15** (1982) L51.

[23] L. S. Miller, S. Howe and W. E. Spear, *Charge transport in solid and liquid Ar, Kr and Xe*, *Phys. Rev.* **166** (1968) 871

[24] E. Aprile, A. Bolotnikov, D. Chen, R. Mukherjee, F. Xu, D. Anderson, V. Peskov, *Performance of CsI photocathodes in liquid Xe, Kr, and Ar*, *Nucl. Instrum. Meth.* A **338** (1994) 328.

[25] E. Aprile, et al., *The XENON100 dark matter experiment*, *Astropart. Phys.* 35 (2012) 573.

[26] C. Rethmeier for the nEXO collaboration, *Characterization of VUV sensitive SiPMs for nEXO*, 2016 *JINST* **11** C03002

[27] L. Baudis, M. Galloway, A. Kish, C. Marentini, and J. Wulf, *Characterisation of Silicon Photomultipliers for Liquid Xenon Detectors*, arXiv:1808.06827.